\documentclass[preprint, amsmath, amssymb, aps, pra]{revtex4-2}

\usepackage{graphicx}
\usepackage[colorlinks=true, urlcolor=blue, citecolor=blue, linkcolor=blue]{hyperref}
\usepackage{cleveref}
\usepackage{newtxtext, newtxmath}

\newcommand{\rmd}{\mathrm{d}}
\newcommand{\micron}{\mu\mathrm{m}}
\renewcommand{\vec}{\boldsymbol}
\DeclareMathOperator{\sign}{sgn}

\begin{document}

\title{Theory uncertainties in predictions of strong-field QED}

\author{T. G. Blackburn}
\email{tom.blackburn@physics.gu.se}
\affiliation{Department of Physics, University of Gothenburg, SE-41296 Gothenburg, Sweden}

\begin{abstract}
Experiments using high-power lasers and relativistic electron beams will soon be capable of precision testing of the theory of strong-field quantum electrodynamics.
The comparison between experiment and theory always occurs via numerical simulations, but the lack of quantitative uncertainty bounds for the latter means that the origin of any discrepancies cannot be identified conclusively.
Here I present a simulation framework that can place meaningful bounds on the theory uncertainties inherent to its operation.
This work shows that the use of the locally constant field approximation leads to a theory uncertainty of a few per cent in the signals of quantum radiation reaction expected in near-term experiments.
\end{abstract}

\maketitle
\newpage

\section{Introduction}

Experiments with high-power lasers are now being used to test the theory of strong-field quantum electrodynamics (QED)~\cite{cole.prx.2018,poder.prx.2018,mirzaie.np.2024,los.2024}.
This theory is nonperturbative in the coupling to the electromagnetic (EM) field~\cite{dipiazza.rmp.2012,fedotov.pr.2023}; it predicts the dominance of high-order processes, such as cascades of high-energy photon emission and electron-positron pair creation~\cite{bell.prl.2008,gonoskov.rmp.2022}, and it is essential for understanding the particle and plasma dynamics in the magnetospheres around astrophysical compact objects~\cite{harding.rpp.2006,philippov.prl.2020}.

For any comparison between theoretical predictions and experimental data to be meaningful, it is essential that both include quantitative uncertainty bounds~\cite{sarri.2025}.
The theoretical predictions are always obtained from numerical simulations, because of the severe difficulty in using the theory directly for high-order processes and experimentally relevant conditions.
Numerical simulations are only able to provide these predictions because they make a series of approximations~\cite{ridgers.jcp.2014,gonoskov.pre.2015}, each of which induces some uncertainty in the final result.
This theory uncertainty, which is needed to judge the agreement or disagreement with experimental data, has not been quantified.

In this work, I present a means by which the theory uncertainty may be quantified for signals of strong-field QED of experimental relevance.
Specifically, we consider the uncertainty that arises from the use of the locally constant field approximation (LCFA) as a demonstration of the general method.
We obtain a meaningful estimate for the error made by this approximation, propagate this error through a simulation of a high-intensity laser interaction, and demonstrate how it affects final-state particle spectra.
As an example, the theory uncertainty in the energy spectrum of electrons that have undergone quantum radiation reaction~\cite{dipiazza.prl.2010,neitz.prl.2013,blackburn.prl.2014,vranic.njp.2016} in an intense laser pulse is a few per cent in magnitude, for current and near-term collision parameters.
This work opens a pathway to determine how well upcoming, higher precision, experiments can test and constrain the theory of strong-field QED, and as well as to determine what precision would be required to make the theory uncertainties visible.

\section{Accuracy of the LCFA emission rate}
\label{sec:Theory}


Numerical simulations depend, in particular, on the locally constant field approximation (LCFA), which is that photon emission or pair creation can be treated as occurring instantaneously; as such, the probability rate is a function only of local parameters, such as the instantaneous quantum parameter $\chi$.
It applies if the formation length of the given process is much smaller than the characteristic scale of variation of the external EM field.
The LCFA arises from a local expansion of the unknown `true' rate of photon emission:
    \begin{equation}
        \frac{\rmd W}{\rmd f} = \frac{\rmd W^{(0)}}{\rmd f} + \frac{\rmd W^{(1)}}{\rmd f} + \cdots,
    \end{equation}
where the leading order term is~\cite{ritus.jslr.1985}:
    \begin{equation}
    \frac{\rmd W^{(0)}}{\rmd f} = \frac{\alpha}{\sqrt{3} \pi \gamma \tau_C} \left[
        \left(1 - f + \frac{1}{1-f} \right) K_{2/3}(\xi)
    \right. \\ \left.
    - \int_\xi^\infty \! K_{1/3}(t) \, \rmd t
    \right].
    \end{equation}
Here $\gamma$ and $\chi_e$ are the electron Lorentz factor and quantum parameter respectively, $f = \epsilon_\gamma / \epsilon_e$ is the ratio of the photon and electron energies, $\xi = 2 f / [3 \chi_e (1 - f)]$ and $K_n$ is a modified Bessel function of the second kind, of order $n$.

The next to leading order (NLO) term, which depends on derivatives of the electromagnetic field, is given by~\cite{ilderton.pra.2019}:
    \begin{align}
    \frac{\rmd W^{(1)}}{\rmd f} &= \frac{\alpha}{\sqrt{3} \pi \gamma \tau_C} \left( \delta_1 \frac{\rmd \mathcal{X}_1}{\rmd f} + \delta_2 \frac{\rmd \mathcal{X}_2}{\rmd f} \right)
    \label{eq:PhotonRateCorr}
    \\
    \frac{\rmd \mathcal{X}_1}{\rmd f} &= \frac{2}{1-f} K_{2/3}(\xi) - \frac{f}{\chi_e (1-f)^2} K_{1/3}(\xi)
    \\
    \frac{\rmd \mathcal{X}_2}{\rmd f} &= \frac{\chi_e(2-f)}{f} K_{1/3}(\xi) - \frac{2-f}{1-f} K_{2/3}(\xi)
    \end{align}
In the limit $f \to 0$, $\frac{\rmd}{\rmd s} \mathcal{X}_2 \simeq 3.864 (\chi_e / f)^{4/3} $ diverges faster than $\frac{\rmd}{\rmd s} \mathcal{X}_1 \simeq 2.817 (\chi_e / f)^{2/3}$.
This divergence at small $f$ means that the correction to the number spectrum is not integrable.

For a plane electromagnetic wave, which is defined by an electric field that is only a function of phase, $\vec{E}(\phi)$, the corrections to the LCFA rates depend on field derivatives in the following two combinations:
    \begin{align}
    \delta_1 &= \frac{3 \vec{E}.\vec{E}'' + \vec{E}' . \vec{E}'}{45 (\vec{E}.\vec{E})^2},
    &
    \delta_2 &= \frac{3 \vec{E}.\vec{E}'' - 4 \vec{E}' . \vec{E}'}{45 (\vec{E}.\vec{E})^2}.
    \label{eq:Delta}
    \end{align}
Here primes denote derivatives with respect to phase $\phi$.

In the paradigmatic case of a circularly polarised electromagnetic wave with normalised amplitude $a_0  = e E_0 / (m c \omega_0)$, where $E_0$ is the electric-field strength and $\omega_0$ is the angular frequency,
    \begin{align}
    \delta_1 &= -\frac{2}{45 a_0^2},
    &
    \delta_2 &= -\frac{7}{45 a_0^2}.
    \label{eq:DeltaCP}
    \end{align}
The normalised amplitude $a_0$ appears here because it contains a factor of the frequency $\omega_0$, which controls the rate of the change of the background field: at fixed intensity (fixed $E_0$), the importance of these corrections increases with $\omega_0^2$.
Expanding \cref{eq:PhotonRateCorr} in powers of $1/a_0$ (at fixed $f$), using \cref{eq:DeltaCP}, we find that the typical size of the correction may be related to the ratio of the energy-resolved photon formation length $L_\gamma$ and the laser wavelength~\cite{blackburn.pra.2019},
    \begin{equation}
    \lim_{a_0 \to \infty} \frac{| \rmd W^{(1)} / \rmd f |}{\rmd W^{(0)} / \rmd f}
        \simeq 0.21 \left( \frac{\chi_e}{a_0^3} \frac{1-f}{f} \right)^{2/3}
        \simeq 9 \left( \frac{L_\gamma}{\lambda} \right)^2,
    \label{eq:PhotonErrora0}
    \end{equation}
which is reasonable, because this ratio characterises the extent to which the external field is indeed `locally constant'.

\section{Implementation of uncertainty tracking}
\label{sec:Implementation}

The proposition of this work is that we treat the absolute value of the lowest-order correction, \cref{eq:PhotonRateCorr}, as an unbiased estimate of the systematic uncertainty associated with the LCFA, i.e. a best guess of the truncation error that results from a local expansion of the QED probability.
On this basis, the `true' rate of photon emission, $W$, is expected to lie somewhere in the region $W^{(0)} - |W^{(1)}| \lesssim W \lesssim W^{(0)} + |W^{(1)}|$.

The nonlinear nature of particle dynamics in the strong-field QED regime means that numerical simulations are needed to study how this uncertainty affects final-state spectra.
An outline of the simulation framework will be given here, with details reserved for a future publication.
In order to propagate a theory uncertainty through a simulation, the rate of photon emission for each individual initial particle is assigned to be:
    \begin{equation}
        \frac{\rmd W}{\rmd f} =
            \frac{\rmd W^{(0)}}{\rmd f} +
            \sign(u) \min \left(
                \left| u \frac{\rmd W^{(1)}}{\rmd f} \right|,
                \frac{\rmd W^{(0)}}{\rmd f}
            \right),
    \label{eq:UncertainRate}
    \end{equation}
where $u$ is a pseudorandom number drawn from an appropriate prior probability distribution, when the particle is initialised.
Here we opt for a standard Gaussian prior, $u \sim \mathcal{N}(0, 1)$, which means that the true rate lies between $W^{(0)} \pm 2 |W^{(1)}|$ at the 95\% confidence level.
The error term is limited to being no larger than the rate itself, in order to avoid negative probabilities being generated~\cite{ilderton.pra.2019}; as a consequence, the largest uncertainty at any point in the spectrum will be 100\%.
Using the absolute value of the correction avoids the appearance of a zero crossing in the systematic uncertainty, which would claim more confidence than is warranted.

Once their $u$ values have been chosen, particles are tracked as they travel through a region of strong field in the usual way: by solving the Lorentz force equation, pseudorandomly sampling the emission spectrum along the classical trajectory, and recoiling the particle when emission occurs~\cite{ridgers.jcp.2014,gonoskov.pre.2015}.
To generate the simulation results in this work, we use the Ptarmigan simulation code~\cite{blackburn.pop.2023}, which has been changed to use the `modified event generator' of Ref.~\citenum{gonoskov.pre.2015} to sample the photon emission spectrum [\cref{eq:UncertainRate}].
As it tracks the incident particles, the code collects, for example, an array of photons with energies $\epsilon_i$ and uncertainty values $u_i$ (inherited from their parent electrons).
The photon energy spectrum, including uncertainty bounds at the $n \sigma$ level, follows as
    \begin{align}
    \frac{\rmd N_\gamma}{\rmd \epsilon} \simeq
        \left. \frac{\rmd N_\gamma}{\rmd \epsilon} \right|_{u = 0}
        \pm n \left. \frac{\partial^2 N_\gamma}{\partial u \partial \epsilon} \right|_{u = 0}.
    \end{align}
Similar logic applies to other final-state spectra.

\begin{figure}
    \includegraphics[width=0.7\linewidth]{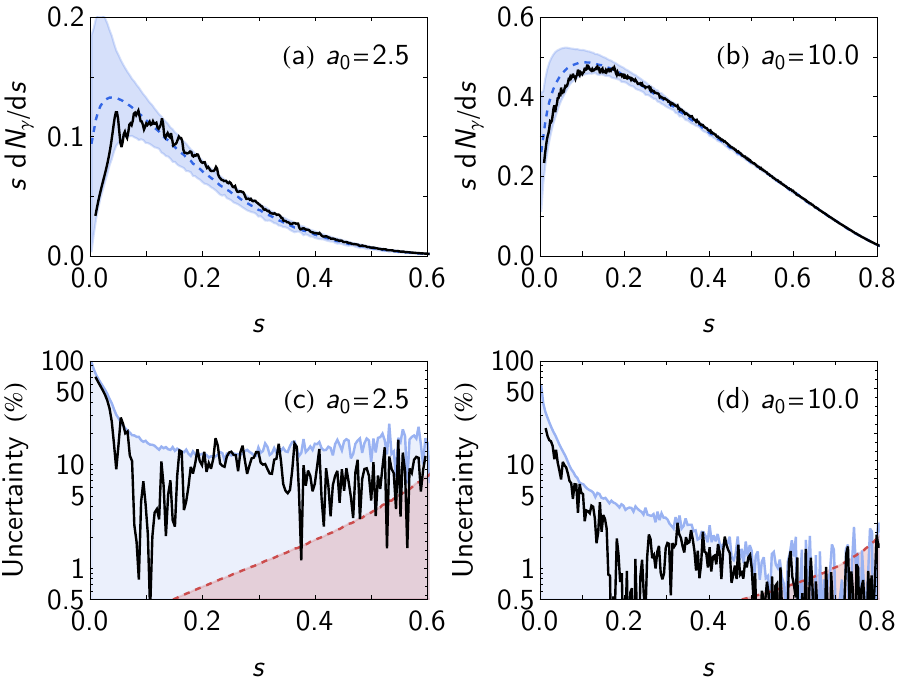}
    \caption{%
        Upper panels: Photon spectra predicted by LCFA-based simulations, including uncertainty bounds at the 95\% confidence level (blue dashed lines and bands), and QED theory (black, solid lines), at fixed electron energy parameter $\eta_e = 0.1$.
        Lower panels: Systematic uncertainty (blue), statistical uncertainty (red) and relative difference between the QED and simulation results (black), all in per cent.
        Theory data reproduced from Ref.~\citenum{blackburn.pop.2023}.
    }
    \label{fig:UncertaintyBenchmark}
    \end{figure}

In order for the systematic uncertainties to be meaningful, they need to be larger than the statistical uncertainties inherent in Monte Carlo simulations.
The latter are calculated as follows, when the code bins the final-state photons in order to generate histograms of the kind shown in \cref{fig:UncertaintyBenchmark}:
if the total weight of photons in a particular bin is $\sum_{i=1}^N w_i$, where $N$ is the number of photons in that bin, the statistical uncertainty (at one standard deviation) is the square root of the variance $\sum_{i=1}^N w_i^2$.

Let us demonstrate that the procedure outlined here leads to reasonable uncertainty bounds by comparing the results of LCFA-based simulations with QED theory predictions for single nonlinear Compton scattering ($e \to e \gamma$).
This kind of benchmarking has previously been reported in, e.g., Refs.~\citenum{harvey.pra.2015,blackburn.pop.2018,dipiazza.pra.2018}.
The geometry we consider is a head-on collision between a relativistic electron and a linearly polarised, plane-wave laser pulse, which has wavelength $0.8~\mu$m and vector potential $e A_x = a_0 m c \sin(\phi) \cos^2(\phi / 32)$ for $|\phi| < 16 \pi$.
The field derivatives needed to calculate \cref{eq:Delta} are defined analytically, given this potential.
The initial value of the electron energy parameter, $\eta_e = 0.1$, is equivalent to an energy of $\epsilon_0 = 8.4$~GeV at this laser wavelength.
The QED result, which is calculated to first-order in the fine-structure constant $\alpha$, does not contain higher-order corrections, such as radiation reaction.
In the simulations we therefore disable recoil on emission, so that we have a consistent basis for comparison~\cite{dipiazza.prl.2010}.

The intensity spectrum of the emitted photons, as predicted by the simulations and QED theory, is shown in \cref{fig:UncertaintyBenchmark}, for two values of $a_0$.
(The theory data have been reproduced from Ref.~\citenum{blackburn.pop.2023}.)
The spectra are given as functions of $s$, the ratio of the photon and electron lightfront momenta, which is approximately equal to the energy ratio $f$ for these collision parameters.
One expects that the simulation results should tend towards the theory as the LCFA becomes more accurate, and indeed this is the case: the relative difference between the two (the black lines in the lower panels) decreases as $a_0$ increases.
However, we are now able to show how the theory and simulation results are consistent with each other at a particular, quantitative uncertainty bound.
In both cases, across the full range of photon momenta, the relative error is smaller than the systematic uncertainty at the 95\% ($2\sigma$) confidence level.
Furthermore, the the bounds are reasonably tight in the way they capture the dependence of the error on $a_0$ and photon momentum.
It should be noted that the comparison is less reliable as at very large photon momenta ($s \to 1$) because the probability of emission is suppressed and there are correspondingly fewer photons per bin: here the statistical uncertainty becomes too large to resolve the systematic uncertainty.
Nevertheless, the comparison confirms that the procedure introduced in this work does provide a meaningful quantitative estimate for the systematic (theory) uncertainty in a simulation result.

\section{Uncertainty in signals of quantum radiation reaction}

Earlier work, reporting experimental measurements of quantum radiation reaction ($e \to e \sum_n \gamma^n$) with high-intensity lasers, has raised the question of whether the mismatch between the experimental results and simulation predictions might arise from the approximations used in simulations~\cite{poder.prx.2018}.
We now examine to what extent the uncertainty caused by the LCFA affects the energy spectrum of an electron beam that collides with a high-intensity laser, radiates, and loses energy.
We consider a 1D interaction between an energetic electron beam, which has a Gaussian energy spectrum (mean energy 2 GeV and 1\% energy spread, rms) and a linearly polarised laser pulse with an intensity parameter $a_0 = 10$, Gaussian temporal envelope (FWHM duration 30~fs) and a wavelength of 0.8 microns.
The energy spectrum of the electrons after the collision is shown in \cref{fig:RadiationReaction}(a), with the relative uncertainty at 2$\sigma$ or 95\% confidence in \cref{fig:RadiationReaction}(b).
The simulation is run using 64 million incident particles, which ensures that the typical statistical uncertainty is less than 0.5\% across a histogram with 100 bins, and that the systematic uncertainties are adequately resolved.
The systematic (theory) uncertainty is approximately 5\% to the left of the peak at $1800$~MeV.
It is larger to the right of this peak and grows with increasing energy, i.e. it is particularly important for electrons that have lost little to no energy.

    \begin{figure}
    \centering
    \includegraphics[width=0.7\linewidth]{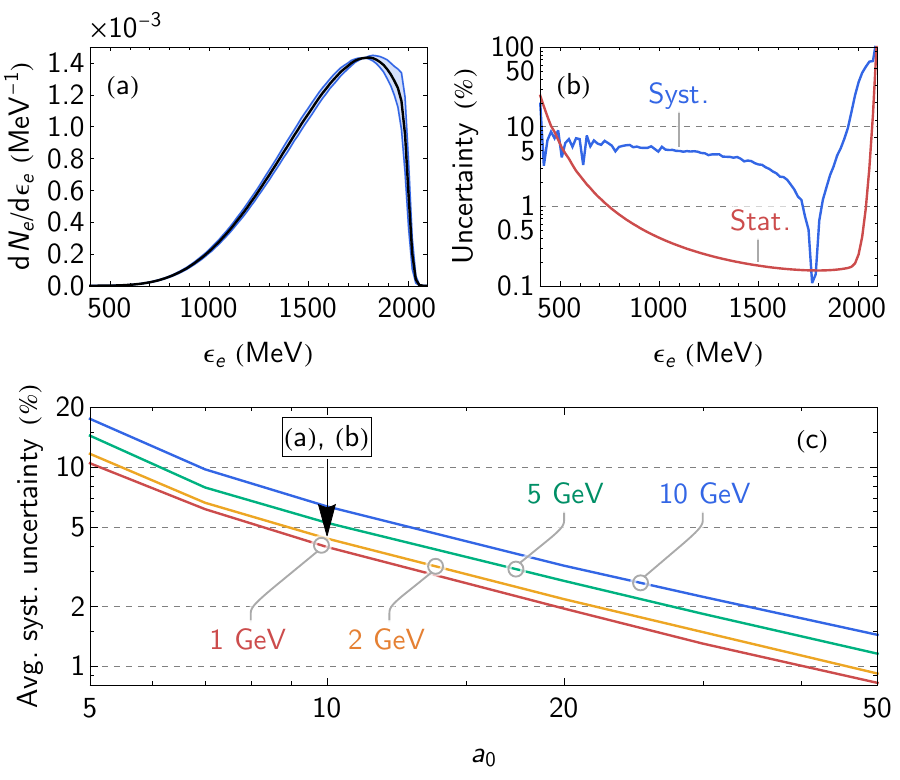}
    \caption{%
        (a) Final-state energy spectrum of a 2-GeV electron beam that has collided with an intense laser pulse ($a_0 = 10$, duration 30 fs), including systematic uncertainties (95\% confidence interval).
        (b) Systematic (blue) and statistical uncertainties (red) for the spectrum in panel (a), both at the 95\% confidence level.
        (c) Average systematic uncertainty in the electron energy spectrum as a function of laser amplitude $a_0$, for four different initial electron energies: (from bottom to top) 1, 2, 5, and 10 GeV.
    }
    \label{fig:RadiationReaction}
    \end{figure}

We may compute an typical systematic uncertainty by averaging the relative uncertainty over the electron spectrum.
The results, for four different initial energies and a range of laser intensities, are shown in \cref{fig:RadiationReaction}(c).
For the parameters of upcoming experiments, the typical scale is a few per cent.
A power-law fit to the data shown in \cref{fig:RadiationReaction}(c) indicates that the average uncertainty scales as $\epsilon_0^p  a_0^q$ ($p = 0.23$, $q = -1.33$).
This should reflect the uncertainty in the radiation power $\delta\mathcal{P}$, which ultimately controls the uncertainty in the electron's energy loss and thus the final energy spectrum.
Given \cref{eq:UncertainRate}, one may show that $\delta\mathcal{P} = \mathcal{P}^{(1)} / \mathcal{P}^{(0)}$ where $\mathcal{P}^{(n)} = \epsilon_e \int_0^1 f \left| \frac{\rmd W^{(n)}}{\rmd f} \right| \rmd f$.
This is proportional to $a_0^{-2}$ and independent of $\chi_e$, if $\chi_e \ll 1$.
The reason for the different scalings is multiple emission: since the electron will radiate several times before escaping the laser pulse, the total uncertainty should scale as $\sqrt{N_\gamma} / a_0^2 \propto a_0^{-3/2}$ (if $\chi_e$ is not too large), which is in reasonable agreement with the simulation results.
On the other hand, the increase of the uncertainty with increasing electron energy is a quantum effect: if $\chi_e \gg 1$, $\delta\mathcal{P} \propto \chi_e^{2/3}/a_0^{2} \propto \epsilon_0^{2/3}$ [cf. \cref{eq:PhotonErrora0}].
The weaker scaling with $\epsilon_0$ obtained from the simulations is a consequence of radiative energy losses.

This paper concludes with an example of how this uncertainty tracking framework can be applied to a more realistic scenario.
In any real experiment, the spatiotemporal structures of the high-intensity laser and electron beam will play a significant role:
signals of strong-field QED are integrated over an effective `peak' $a_0$, which varies across the transverse plane~\cite{amaro.njp.2021}, as are any uncertainties.
The framework presented here may be extended to cover the case of a focusing laser pulse by means of a high-energy approximation~\cite{ilderton.pra.2019}, such that \cref{eq:Delta} becomes
    \begin{align}
    \delta_1 &= - \frac{3 \dot{u}.\dddot{u} + \ddot{u} . \ddot{u}}{45 (\ddot{u}.\ddot{u})^2},
    &
    \delta_2 &= \frac{4 \ddot{u} . \ddot{u} - 3 \dot{u}.\dddot{u}}{45 (\ddot{u}.\ddot{u})^2},
    \label{eq:Delta3D}
    \end{align}
where $\dot{u} = d u / dt$ is the (laboratory frame) time derivative of the four-momentum.
The momentum derivatives may be obtained approximately by storing past values of $\dot{u}_\mu \propto F_{\mu\nu} u^\nu$ (the force exerted by the external field) during the tracking process and computing finite differences, e.g., $\ddot{u} \simeq \Delta \dot{u} / \Delta t$, where $\Delta t$ is the simulation timestep.
A crosscheck for the case of a pulsed plane wave has established that using \cref{eq:Delta3D} and an appropriately small timestep, $\Delta t \lesssim 1/(2 \omega_0)$, does reproduce the uncertainty calculated directly from \cref{eq:Delta}.

    \begin{figure}
    \centering
    \includegraphics[width=0.7\linewidth]{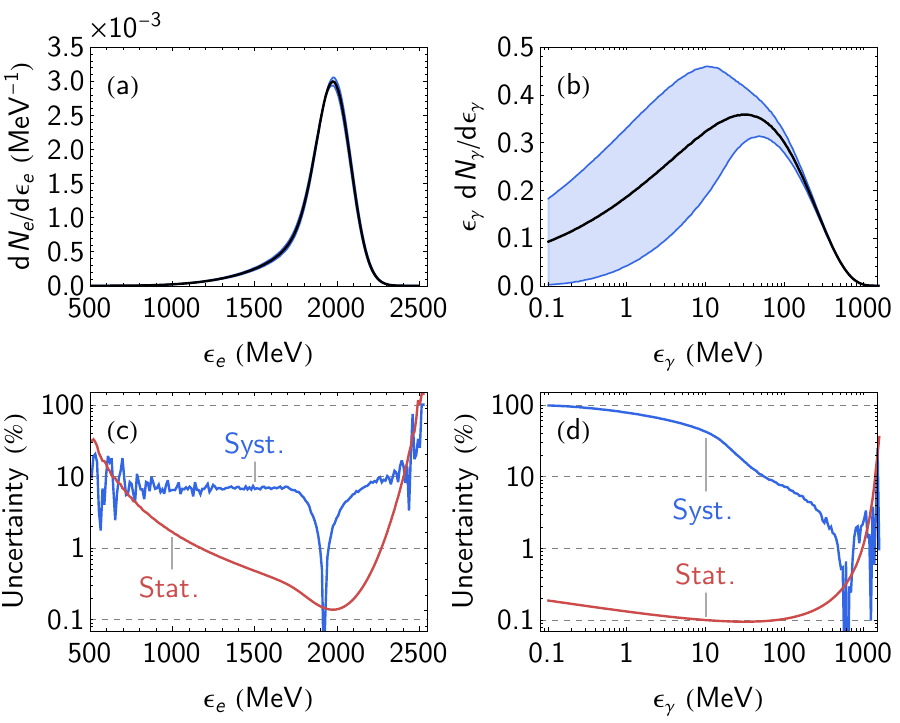}
    \caption{%
        (a) Electron and (b) photon energy spectra for a 2-GeV electron beam that has collided with an intense laser pulse ($a_0 = 10$, duration 30 fs), including systematic uncertainties (95\% confidence interval).
        (b) and (c): Systematic (blue) and statistical uncertainties (red) for the spectra, both at the 95\% confidence level.
    }
    \label{fig:RadiationReaction}
    \end{figure}

We consider parameters that would be reasonable for a near-term experiment and therefore simulate the collision between an electron beam with Gaussian energy spectrum (mean 2 GeV, standard deviation 100 MeV) and charge distribution (radius $3.0~\micron$ and length $6.0~\micron$), and a laser pulse that has focal spot size $w_0 = 3.0~\micron$, peak $a_0 = 10$, wavelength $0.8~\micron$ and duration $30$~fs.
The energy spectrum of the electrons post-collision, and that of the photons they emit, is shown in \cref{fig:RadiationReaction}(a) and (b).
The relative uncertainties, both systematic and statistical, are shown below in panels (c) and (d).
A typical value of the theory uncertainty in the electron spectrum is 8\%; the average value, spectrally weighted, is 4\% (both at the $2\sigma$ level).
In the context of recent experiments~\cite{cole.prx.2018,poder.prx.2018,mirzaie.np.2024,los.2024}, this uncertainty is a smaller contribution than fluctuations in the collision parameters.
However, the former are also smaller than the differences between the electron spectra predicted by classical or modified classical models of radiation reaction.
Thus violation of the LCFA would not, by itself, prevent an experiment discriminating between these models.
The uncertainty in the photon spectrum is larger than 5\% for all energies below 100~MeV, because there are contributions from electrons that have hit the fringes of the laser pulse, where the local peak $a_0$ is smaller and the LCFA less reliable.
Use of simulated data to assist reconstruction of photon spectra in the few MeV range should be approached with caution, as there are significant theory uncertainties here.

\section{Summary and outlook}

In conclusion, this work takes a crucial first step towards the important goal of placing a quantitative uncertainty bound on the simulation predictions of strong-field QED interactions~\cite{sarri.2025}.
Note that these theory uncertainties are distinct from uncertainties that arise from fluctuations in initial conditions (e.g. laser intensity, particle energies, or beam alignment).
The latter are not inherent to the operation of a numerical simulation and they are readily estimated by the use of existing codes~\cite{cole.prx.2018,poder.prx.2018,mirzaie.np.2024}.
Using the post-LCFA result from \citet{ilderton.pra.2019} enables us to establish a meaningful value for the error induced by the LCFA, which is verified by benchmarking against results for single nonlinear Compton scattering.
Extending this to the case of quantum radiation reaction, we find that the theory uncertainty in the electron final energy spectrum is several per cent for the parameters of upcoming experiments.
The uncertainty shrinks with $a_0$, but grows with electron energy and the number of strong-field QED interactions per particle.

Future work will include the LCFA-induced uncertainty in the electron-positron pair creation rate~\cite{king.pra.2020}, in order to study the theory uncertainty in simulations of pair showers~\cite{,blackburn.pra.2017,lobet.prab.2017,pouyez.pre.2024} and avalanches~\cite{bell.prl.2008,gelfer.pra.2015,grismayer.pre.2017,gonoskov.prx.2017,mercuribaron.prx.2025}.
Furthermore, the method used here could also be applied to determine the uncertainty associated with the cascade approximation, which is assumed in order for high-order processes to be modelled as the incoherent product of first-order processes~\cite{fedotov.pr.2023}.
In this case, the equivalent of the neglected (NLO) term in an expansion of the probability of a second order process, such as trident pair production ($e \to e^- e^+ e$), would be the so-called `one-step' contribution~\cite{ilderton.prl.2011,king.prd.2013,dinu.prd.2018,mackenroth.prd.2018}.

\begin{acknowledgments}
The author thanks A. Ilderton, B. King and D. Seipt for helpful discussions and S. Tang for providing the theory data from Ref.~\citenum{blackburn.pop.2023}.
\end{acknowledgments}

\clearpage

\bibliography{references}

\end{document}